\def\cm{{\rm\thinspace cm}}
\def\dyn{{\rm\thinspace dyn}}
\def\erg{{\rm\thinspace erg}}

\def\keV{{\rm\thinspace keV}}

\def\kpc{{\rm\thinspace kpc}}

\def\Msun{\hbox{$\rm\thinspace M_{\odot}$}}

\def\s{{\rm\thinspace s}}
\def\yr{{\rm\thinspace yr}}

\def\dynpcmsq{\hbox{$\dyn\cm^{-2}\,$}}

\def\ergps{\hbox{$\erg\s^{-1}\,$}}

\def\Msunpyr{\hbox{$\Msun\yr^{-1}\,$}}

\def\mdot{\hbox{$\dot{m}$}}
\def\Mdot{\hbox{$\dot{M}$}}
\def\spose#1{\hbox to 0pt{#1\hss}}
\def\approxlt{\mathrel{\spose{\lower 3pt\hbox{$\sim$}}
        \raise 2.0pt\hbox{$<$}}}
\def\approxgt{\mathrel{\spose{\lower 3pt\hbox{$\sim$}}
        \raise 2.0pt\hbox{$>$}}}

\documentclass[]{article}
\usepackage{emulateapj}
\usepackage{psfig}
\begin{document}

\title
{LIMITS ON THE ACCRETION RATES ONTO MASSIVE BLACK HOLES IN NEARBY GALAXIES}

\author{T.~Di~Matteo\footnote{{\em Chandra} Fellow}}\affil{Harvard-Smithsonian Center 
for Astrophysics,60 Garden St., Cambridge, MA 02138; 
tdimatteo@cfa.harvard.edu}
\author{C.~L.~Carilli}\affil{NRAO, P.O. Box O, Socorro, NM, 87801; ccarilli@aoc.nrao.edu}  
\author{and}
\author{A.~C.~Fabian}
\affil {Institute of Astronomy, Madingley Road, Cambridge, CB3 OHA, UK; acf@ast.cam.ac.uk }

\medskip

\setcounter{footnote}{0}

\begin{abstract}

The radio emission from supermassive black holes in nearby early-type
galaxies can be used to test possible explanations for their low
luminosities.  We calculate the expected contribution from thermal
synchrotron emission from hot accretion flows to the high radio
frequency observations of NGC 2300, NGC 1399, NGC 4278 and NGC
4594. We find that, in all cases, and in accordance with our previous
findings, hot flows accreting close to their Bondi rates overestimate
significantly the observed fluxes. This implies that simply assuming a
low radiative efficiency for the accreting gas is not enough to
explain their low luminosities. Smaller central densities and
accretion rates, as expected in the presence of strong mass loss or
convection in the flows, can help reconcile the models with
observations. We also show that a significant contribution to the
high-frequency radio spectra can be due to non-thermal synchrotron
emission from the small scale radio jets observed in these systems,
allowing for even lower accretion rates in the inflows. We suggest
that these outflows or jets may dump significant energy into the
surronding medium close to the accretion radius and so reduce the
accretion rates onto these systems.  We discuss the relationship
between the radio flux and black hole mass for the observed sample and
its potential importance for probing accretion models.

\end{abstract}


\section{Introduction}

Most nearby galaxies exhibit little or no nuclear activity. However
dynamical arguments based on the observed stellar and gas
distributions firmly imply the presence of supermassive compact
objects in their cores (Magorrian et al.~1998; Richstone et al.~1999;
van der Marel 1999). These studies show that virtually all early-type
galaxies host black holes with masses in the range $10^8-$ to a few
$10^9$ \Msun.

The central black holes in nearby early--type galaxies are probably the
remnants of QSO phenomena (McLure et al.~1999). Unlike the giant
ellipticals at high redshifts which host radio galaxies and radio-loud
quasars, they only display low-luminosity radio cores (Sadler, Jenkins
\& Kotanji 1989; Wrobel \& Heershen 1991). However, the black holes in the
centers of nearby early--type galaxies have enough fuel so that they
should still exhibit quasar--like activity. X-ray studies show that
they possess extensive hot-gaseous halos which should accrete onto the
central black holes and give rise to far more activity than is
observed. The Bondi accretion rates for the typical temperatures and
densities of their interstellar medium (ISM) are typically estimated
to be $\sim 0.01-0.1$ \Msunpyr, implying luminosities of $10^{45-46}
\ergps$, for standard accretion disks radiative efficiencies of $\sim
10\%$ (see e.g. Di Matteo et al. 2000 for details). Because of the
lack of any such activity it has been suggested (Fabian \& Rees 1995;
Reynolds et al. 1996; Mahadevan 1997; Di Matteo \& Fabian 1997, Di
Matteo et al. 1999, 2000 and references therein) that accretion in the
nuclei of ellipticals occurs at low radiative efficiency as predicted
in advection dominated accretion flow (ADAF) models (e.g. Rees et
al. 1982; see Narayan, Mahadevan \& Quataert 1998 for a review). In
these models, thermal synchrotron emission is predicted to give a
strong contribution to the radio emission.

Because the only obvious sign of activity from these supermassive
black holes is their radio emission, studies in this band provide a
useful tool for constraining accretion models for these systems. In
order to discriminate between a potential accretion flow component and
emission from the more extended, scaled-down jets, also common around
the supermassive black holes in early-type galaxies, it is crucial to
examine their core radio emission component at high resolution and
high radio frequencies. In previous work we have shown that the
high-frequency radio emission from the cores of three elliptical
galaxies is strongly suppressed with respect to the standard ADAF
model predictions for black holes accreting at close to the Bondi
rates (Di Matteo et al. 1999, hereafter DM99). We also showed that
radio emission can provide powerful constraints on ADAF models for
the cores of ellipticals, implying that the low-radiative efficiency
of an ADAF in not enough to explain their low luminosities.  and that
the accretion rates onto the black holes ought to be smaller than the
expected Bondi rates. We have examined how the required suppression
can be obtained if strong mass loss is present in hot accretion flows
and/or if matter is fed at much lower rates than those expected
(DM99). This is in agreement with the proposal that a direct
consequence of the dynamics of hot quasi-spherical flows is the
development of strong outflows (as emphasized by Begelman \& Blandford
1999; Igumenshchev \& Abramowicz 1999; Stone, Pringle \& Begelman
2000) leading to suppressed density profiles $\rho \propto
r^{-3/2+p}$ for $0<p<1$ in the flows. For certain regimes, convection
might also become important in these hot flows, leading to virtually
no accretion and $\rho
\propto r^{-1/2}$ (Quataert \& Gruzinov 2000; Narayan, Igumenshchev \&
Abramowicz 2000).

To test how representative our previous results are and how they can
constrain the properties of accretion flows in the low luminosisity
nuclei of nearby early-type galaxies, here we extend our analysis to a
larger sample of objects observed at high radio frequencies. We
present accretion models based on Very Large Array (VLA) observations
at 8, 22 and 43 GHz of four further galaxies, NGC 1399, NGC 2300, NGC
4594, and NGC 4278. We have also re-observed NGC 4649, a Virgo
elliptical from our initial sample to test for variability. With the
three ellipticals previously studied, we now have seven objects with
radio spectra which can be modeled from $\sim 5$ GHz to $43$ GHz.

In the next section we briefly summarize the observations (\S 2) and
in \S3 discuss possible accretion and jet models for the observed
radio emission (\S 3). In \S 4, we derive constraints for hot
accretion models. In \S 5 we discuss how the presence of small-scale
radio jets in these systems should heat the ISM close to the accretion
radii with a consequent reduction of the accretion rates onto the
black holes (independent of any specific consequences of the dynamical
properties of hot accretion flows themselves). We also discuss the
relationship between radio flux and black hole mass in these objects
and show that it can provide a useful tool for constraining accretion
flow properties.

\section{The VLA data and the spectrum of the core emission}

Radio continuum surveys of elliptical and S0 galaxies have indicated
that the sources in radio--quiet galaxies tend to show compact
components which have relatively flat or slowly rising radio spectra
(with a typical spectral index $\alpha \sim 0.3-0.4$, where flux,
$F_{\nu} \propto \nu^{-\alpha}$) suggesting that the radio emission
from early-type galaxies is, in general, of nuclear origin (Slee et
al. 1994; Wrobel 1991).

Our primary goal in obtaining high-frequency, high resolution radio
measurements of early-type galaxy radio cores is the determination of
their radio spectral energy distributions. This is important because
only at high radio frequencies is the inverted-spectrum due to
synchrotron radiation from an ADAF (for accretion rates close to the
Bondi values) expected to dominate over the flatter spectral component
of their compact radio cores. The observation of such an inverted
component is crucial for determining the presence of hot-thermal
accretion flows around the supermassive black holes in these galaxies
and hence for understanding the dominant mode of accretion in these
extremely underluminous systems.

Here we use the VLA to extend the sample of objects so far observed at
high radio frequencies and present results for NGC 1399, NGC 2300,
NGC 4594 and NGC 4278. Following DM99, in the attempt to resolve the
intrinsic spectra of the central point sources of these compact radio
cores we obtain flux measurements at 8.4, 22 and 43 GHz. In order to
check for core flux variability we have also re-observed NGC 4649 which was
the only unresolved point-like source in DM99.

Both NGC 1399 and NGC 2300 have been fairly extensively studied at low
radio frequencies and in the X-ray band. NGC 1399 is a giant elliptical at
the center of the Fornax Cluster at a distance of $\sim 30$ Mpc. VLA
imaging (6 and 20 cm observations by Killeen, Bickenell \& Ekers 1988;
and references therein) of NGC 1399 shows that the source is extended
with antiparallel jets ($\approxlt 4$ arc min) with small diffuse
lobes mostly confined within the extent of the galaxy.

NGC 2300 is an S$\emptyset$ galaxy at a distance of $\sim$ 40
Mpc. Previous VLA observations (Fabbiano, Gioia \& Trinchieri 1989)
indicate a total flux density of 0.7 mJy at 5 GHz for the source with
only upper limits on the flux at 1.4 and 2.4 GHz (Hummel 1980). No
radio maps are available.

NGC 4594 is an Sa galaxy at a distance of $\sim 10$ Mpc with a central
black hole mass of $M_{\rm BH} \approx 2-4 \times 10^9 \Msun$
(Kormendy et al.~1996). It has a prominent bulge with X-ray properties
similar to those of E and S0 galaxies (e.g.~Fabbiano \& Juda
1997). VLA high-resolution data between 0.6 and 15 GHz define a
compact flat-spectrum core with $\alpha$ ranging from 0.2 to 0.4
(Hummel, Hulst \& Dickey 1984).

NGC 4278 is an elliptical radio galaxy at a distance of $10-13$ Mpc
(Jacoby et al. 1996; Forbes et al. 1996). Previous 1.4, 5, 15 GHz
high-resolution radio measurements of the core spectrum were reported
by Wrobel (1991).

\subsection{VLA measurements}
Observatons of NGC 1399, NGC 2300, NGC 4649 were obtained in December
1998, and those of NGC 4594 and NGC 4278 in February 2000 with the VLA
at 8.4, 22 and 43 GHz. As an example, in Figure 1 we show the
resulting high-resolution radio images for NGC 1399. Data for all the
objects are shown in Table 1. In order to obtain the best limits on
the core flux for the extended sources, the flux densities, where
possible, were derived from images convolved with the resolution of
the 8.4 GHz image at all frequencies (seventh column in
Table~\ref{t:vla}). The observations were all made in the VLA C
configuration with a maximum baseline of 3 km. The flux scale was set
using observations of 3C 286.

{\it NGC 1399}: The source is extended at 8 GHz, with prominent
twin jets. The companion source to the north east (Fig.~1) has a  11.2 mJy
total flux density at 8 GHz, a peak flux of 4.4 mJy at RA 03 38 43.039
and DEC -35 23 40.95 (J2000). It has been identified with an elliptical
galaxy (possibly a Seyfert 2) in a cluster behind the Fornax cluster
(Killeen et al. 1988; Carter \& Malin 1983). The core spectrum of NGC
1399 shows evidence for a slowly rising, high frequency component
turning over at $\sim 30$ GHz (Fig.~2).

{\it NGC 2300}: The core of the galaxy is detected with flat/slowly rising
spectrum up to 22 GHz, but only at a 4.5 $\sigma$ significance (see
contour plots in Fig.~1 and spectrum in Fig.~2). It is not detected at
43 GHz with a 4 $\sigma$ limit of 1.6 mJy.

{\it NGC 4649}: In agreement with our previous observations at these
frequencies, NGC 4649 is a core dominated source. At 8 GHz, some
faint, fuzzy emission to the northeast and a possible jet with a knot
at larger distance is observed. The knot has a peak flux of 0.3 mJy
and the total flux density of the fuzzy jet emission is 0.77 mJy. The
source variability, when compared to our previous observations is less
than 15 per cent (see DM99).

{\it NGC 4594}: 8, 22 and 43 GHz measurements indicate a fairly steep
spectrum for this source (Table 1; Fig.~2). Combining with previous
lower frequency observations this implies a spectral turnover at $\sim
10$ GHz.

{\it NGC 4278}: The steep spectral slope (Fig.~2) is likely to be
the result of the dominance of emission from the extended jet
structures present in this system. This component seems to dominate
throughout the radio band and no sharp spectral break is present.

\section{Models}

\subsection{ADAFs with/without outflows: thermal models}

In a hot accretion flow around a supermassive black hole, the majority
of the observable emission arises in the radio and X--ray bands. In
the radio band the emission results from synchrotron radiation due to
the thermal relativistic electrons moving in the near equipartition
magnetic field in the inner parts of the accretion flow. The X-ray
emission is due to either bremsstrahlung or inverse Compton scattering
of the soft synchrotron photons.

In the thermal plasma of an ADAF the self-absorbed synchrotron
emission rises with frequency, $\nu$, roughly as $\sim \nu^{0.3-0.4}$
in the Rayleigh-Jeans limit, up to a critical turnover frequency above
which the emission becomes optically thin and drops abruptly. The peak
emission always arises from close to the black hole and reflects the
properties of the accreting gas within a few Schwarzschild radii. The
spectral models with and without outflows and self-consistent
temperature profiles which we use here are described in detail in DM00
(and references therein).

The predicted spectrum from an ADAF depends on several microphysics
parameters, notably the ratio of gas to magnetic pressure $\beta$, the
viscosity parameter $\alpha$, and the fraction of the turbulent energy
in the plasma which heats the electrons, $\delta$. Here we fix $\alpha
= 0.1$, $\beta = 10$, and $\delta = 0.01$. The two major parameters
although, are the accretion rate $\Mdot$ and the black hole mass
$M_{BH}$.

Models are calculated using the black hole masses reported by Richstone et
al. (1998) and Magorrian et al. (1998; second column, Table
1). Following our earlier work, the accretion rates in the flows
have been determined from Bondi accretion theory e.g. $\Mdot_{\rm
Bondi} \propto M_{\rm BH}^2 n(ISM)/c_{\rm S}(ISM)$ taking ISM
densities $n(ISM)$ at a distance of 1 kpc from the centers and sound
speeds $c_{\rm S}(ISM)$, determined from deprojection analysis of the
ROSAT HRI data and spectral analysis of ASCA data respectively, for
both NGC 1399 (see also Section 2.1 and Table 1 in DM00) and NGC
2300. Because the accretion radii ($R_{\rm A}= GM_{BH}/c_{\rm s}^2$)
of these systems are typically at $\sim 0.1 \kpc$ (with the ISM
temperature $\sim 1 \keV$) and the density profile of the hot gas
scales as $n (ISM)
\propto r^{-1}$ we extrapolate for the density at $R \sim  R_{\rm A}$.
This implies a value of $\sim 0.1\Msunpyr$ and $\sim 1 \Msunpyr$ for
NGC  2300 and NGC 1399 respectively.  For NGC 4594 and NGC 4278 no
deprojection analysis is available.  The X-ray temperatures of their
ISM are estimated to be $kT \sim 0.45$ and $ 0.5 \keV$ for NGC 4594
(Fabbiano \& Juda 1997) and NGC 4278 (Brown
\& Bregman 1998) respectively. These values are consistent with those
of the soft X-ray emission reported for a sample of elliptical
galaxies observed with ASCA. For such systems, a deprojection analysis
shows that the ISM typically has a density $n(ISM) \sim 0.1-0.5$
cm$^{-3}$ (e.g.~Buote \& Fabian 1998).  Given the black hole masses in
Table 1, the Bondi accretion rates for NGC 4594 and 4278 are
therefore expected to be $\Mdot \sim 0.01-0.1 \Msunpyr$.

We measure radii in the flow in Schwarzschild units: $R = rR_{\rm S}$,
where $R_{\rm S} = 2GM_{\rm BH}/c^2$.  We measure black hole masses in solar
units and accretion rates in Eddington units: $M_{\rm BH} = m \Msun$ and
$\Mdot= \mdot \Mdot_{\rm Edd}$. We take $\Mdot_{\rm Edd} = 10L_{\rm
Edd}/c^2 = 2.2 \times 10^{-8} m \Msun$ yr$^{-1}$, i.e., with a
canonical 10\% efficiency.

We model the different density profiles (flatter than in the pure
Bondi inflow with $\rho \propto r^{-3/2}$) by adopting a mass inflow
rate which satisfies: $\mdot \propto r^p$. This relation is supported
by recent numerical and analytical work (Stone et al. 1999;
Igumenshchev et al. 1999; Blandford \& Begelman 1999) which has shown
that mass loss via winds in hot accretion flows may be both
dynamically crucial and quite substantial. In particular, we adopt $
\rho \propto r^{-3/2-p}$. We note that for  $p=1$ and $\alpha
\approxlt 0.1$, this density profile can also correspond to the case of
 strong convection; Narayan et al. 2000; Quataert \& Gruzinov
2000). 

\subsection{Jets: non-thermal models}
Compact radio cores or scaled down radio jets are also synchrotron
emitters. The spectra of compact radio cores are generally flatter
than those of extended sources and are often modeled by non-thermal
synchrotron radiation from power law distributions of
particles. Because of the relative compactness of these sources,
synchrotron self-absorption is also often important in this regime.
At frequencies low enough for the source to be optically thick, the
non-thermal synchrotron spectrum rises as $\propto
\nu^{5/2}$ and at frequencies high enough to be optically thin it falls
$\propto
\nu^{-\alpha}$, where $\alpha = (\xi -1)/2$ for an electron
distribution function $dn/d\gamma \propto \gamma^{-\xi}$, where
$\gamma$ is the Lorentz factor (e.g.~Rybicki
\& Lightman 1979). In between the spectrum has a well-defined peak
near a frequency, which, according to standard synchrotron formalism
we write as,
\begin{equation}
\nu_{t} = 2.8 \times 10^6 B \left(\frac{(\xi - 1) f(\xi) e \tau_T}{B \sigma_{T}} \right)^{2/(\xi +4)},
\end{equation}
for which the synchrotron optical depth is unity. Here, $\tau_T$ is
the electron scattering optical depth across the radius of the source,
$f(\xi)$ is given by Blumenthal and Gould (1970) and for $\xi \sim
2-3$, $f(\xi)
\sim 10$.  For compact sources the observed brightness temperature 
at $\nu_t$ can be derived by assuming that the surface brightness 
is equal to the source function e.g. 
\begin{equation} 
T_{B} = 9.3 \times 10^9 \frac{g(\xi)}{f(\xi)} \left(\frac{(\xi-1)f(\xi)e\tau_T}{B\sigma_T} \right)^{1/(\xi+4)},
\end{equation}
where $f(\xi)$ generally $\sim O(1)$ is also given by Blumenthal and
Gould (1970). The brightness temperature gives the energy and hence a
typical $\gamma$ for the electrons radiating at a frequency $\nu_{t}$;
e.g. $\gamma \sim kT_B/m_ec^2$.  Spectra of radio cores do not usually
rise as sharply as $\nu^{5/2}$ but their flatter spectral shape can be
easily attributed to inhomogeneity. We model the radio spectrum with
regions of different optical depth and magnetic field strengths and
combine them to reproduce a spectrum that matches the one observed for
the different objects over the observed frequencies. A jet provides an
extremely natural context for inhomogeneous models. In a jet, because
density and magnetic field strength plausibly decline outwards, a
number of localized optically thick regions are likely to give rise to
the observed emission. From the flux at a frequency $\nu_{t}$, as given
by our VLA observations and using Eq.~(1) and~(2), we construct simple
models for the spectra and deduce approximate optical depths and
magnetic field strengths for the given sources.

\section{Results: Comparison with the data}

Figure 2 shows our VLA data points and the predicted ADAF
self-absorbed synchrotron emission. We note that, as discussed \S 2.1
and shown in Figure 1, all of the contribution from the weak jets in
these systems cannot be ignored and the measured radio fluxes should
only be considered as upper limits to the emission from a hot
accretion flow in their cores.

Regardless of that and in agreement with our previous results (DM99)
we find that in all cases the canonical ADAF model (solid lines in
Fig.~2) greatly overestimates the total flux contribution at most
frequencies. In addition, although the pure inflow models require
the synchrotron emission to peak at $\nu
\sim 10^{12}$ Hz, all the objects observed here and in DM99 have
spectral energy distributions peaked at $\sim 10-30$ GHz, i.e. at
energies much lower than predicted by ADAF models.

The emission at the self-absorbed synchrotron peak arises from the
inner regions of an ADAF and scales (in the Rayleigh-Jeans limit) as
\begin{equation}
\nu L_{\nu} \propto \nu_{\rm s}^2 T_{\rm e} \propto B^3 T_{\rm e}^7
\propto \beta^{-3/2}\mdot^{3/2} T^{7},
\end{equation}
where $\mdot$ is the accretion rate in the innermost regions of the
 flow. As discussed in DM99 thermal synchrotron emission cannot be
 suppressed significantly by decreasing the magnetic field strength,
 B. A decrease in B leads to an increase in the flow temperature and
 hence an increase in synchrotron cooling itself (unless $B$ is
 reduced to values below 0.1\% equipartion; see Eq.~3 and DM99). In
 addition, the lack of any significant variability of the high
 frequency emission from the most point-like source in our sample, NGC
 4649 (Table 1), and the faint high frequency radio fluxes
 systematically observed in all seven sources, excludes the
 possibility that significant variability may be the cause of the lack
 of the high energy radio flux.

However, if the accretion rates in the inner regions of an ADAF (where
all of the high energy emission is produced; see Eq.~(3)) are
decreased with respect to the Bondi values discussed in \S 3.1, the
ADAF models can be brought to agreement with the data.  There are two
ways this can be achieved (see also DM99). The presence of mass loss
implies $\mdot \propto r$ which can lead to significant suppression of
the central densities and accommodate the observed lack of a high
energy synchrotron component (in particular because the electron
temperature $T_{\rm e}$ also decreases as $p$ increases giving rise to
a steep decrease in the synchrotron component; Eq.~3). An alternative
possibility is that the gas is not fed to flows at the expected Bondi
rates (see \S5.1 and Di Matteo \& Fabian 2000) and $\mdot$ is much
smaller than estimated in \S 3.1 or in previous work (but possibly
consistent with values deduced directly from the lower frequencies radio
flux; e.g.~Wrobel \& Herrnstein 2000).
 
Spectral models which include strong mass loss are shown by the long
dashed lines in Fig.~2. For these models we adopt an accretion rate
which satisfies $\mdot = \mdot_{\rm Bondi}(r/r_{\rm out})^p$. For NGC
2300 the dashed line is for $\mdot_{\rm Bondi} = 2.6
\times 10^{-3}$, $ r_{\rm out}=300$ and $p=0.68$. For NGC 1399 we used
$\mdot_{\rm Bondi} = 10^{-2}$, $p=0.78$ and $r_{\rm out}=220$.  For
NGC 4594 $\mdot_{\rm Bondi} = 3\times 10^{-3}$, $p=0.5$ and $r_{\rm
out}=300$.  We use $m=4 \times 10^9$ to fit the low frequency radio
data.  Finally for NGC 4278 $\mdot_{\rm Bondi} = 2\times 10^{-3}$,
$p=0.4$ and $r_{\rm out}=200$. We note that similar fits (but with
slightly steeper rising spectra) are obtained with $p \sim 1$ and
$R_{\rm out} \sim 10^4$, corresponding to the full extent of the flow.

In all of the above cases, the values of $p$ and $r_{\rm out}$ imply
that only a few percent of the Bondi mass accretion rates are actually
accreted onto the black holes and that most of the mass is lost in the
outflows

In Figure~2 we also show models for which $\mdot$ is again constant
but $\ll \mdot_{\rm Bondi}$ (short dashed lines) so as to agree the
radio limits. We find that values of $\mdot \approxlt $ a few $\times
10^{-5}$ ($\approxlt 10^{-2} \mdot_{\rm Bondi}$) are consistent
with the radio measurements.  We note that for these low accretion
rates the temperature profiles steepen causing the slope of the
synchrotron component also to steepen. As a result, the radio fluxes
are fitted less well by these models and definitely require an
additional more extended component (most likely the jets) to account
for a significant fraction of the observed fluxes.

In Figure~3 we also model the radio spectra of the four galaxies by
non-thermal synchrotron emission from compact 'jet-like' regions as
described in \S 3.2. In the ADAF models, we have assumed that there is
no contribution to the observed emission from either the outflows
(which are considered non-radiative) or from compact regions in the
scaled-down radio jets also present in these sources.  Here we test
whether non-thermal synchrotron models can also explain the radio
spectral energy distributions and/or whether the physical parameters
they require may favor the hot-accretion flow interpretation. Figure 3
shows that if the accretion rates in an ADAF are indeed stifled,
either in the inner regions or throughout the hot flow such as is
implied by our findings above (or if indeed these systems are
accreting via standard thin disks at even lower accretion rates), the
radio data can be easily explained by standard non-thermal models for
compact radio sources.

If we adopt the hypothesis of inhomogeneity (\S 3.2), the observed
spectral breaks in the different sources and the general spectral
shapes can be modeled simply by the emission from different compact
regions emitting self-absorbed non-thermal synchrotron emission.  In
order to reproduce the steep fall off above $10-30$ GHz for most
objects or indeed the typically steep, non-thermal spectrum of NGC
4278 we take $\xi = 3$.  To infer rough values of optical depths and
magnetic field strengths in the different regions we designate the
frequencies $\nu_t$ and use the brightness temperatures and (upper
limits for) the sizes implied by our VLA measurements at the
appropriate frequencies. We then use Equations (1) and (2) and the
models shown in Figure 3 and infer $B
\sim 7 \times 10^{-4} \nu_{t \rm {GHz}} T_{B 12}^{-2}$ G and $\tau_T \sim
0.4 T_{B 12}^{5} \nu_{t {\rm GHz}}$, which are typical values for core
dominated, synchrotron self-absorbed sources. We note that the models
in Figure 3 are not unique (different particle distributions and
different regions can be used) but they provide fairly solid estimates
for the ranges of parameters that characterize the sources, and imply
that the electrons we see directly have $ \gamma \sim $ a few
100. Similar models have been constructed to model the radio spectral
energy distribution of Sgr A$*$ and other galactic radio cores
(e.g.~Falcke \& Biermann 1999; Beckert \& Duschl 1997).

\section{DISCUSSION} 

We have found that all of the seven galactic radio cores observed here
and in DM99 show severe discrepancies with the ADAF model
predictions. Independent of its actual origin, the observed radio flux
is always much less than that expected from pure inflow ADAFs
accreting close to their Bondi accretion rates. Even in those objects
where a spectral turnover is observed, as expected from thermal
synchrotron emission from these hot accretion flows, it is usually at
frequencies $\nu \sim 20-30$ GHz, much lower than those predicted from
ADAF models. We have examined the constraints imposed on ADAF models
by the high-frequency radio observations.

We have shown that, if indeed these supermassive black holes accrete
via hot accretion flows, the suppression of the synchrotron component
(in agreement with the radio observations), implies that the accretion
rates onto the central black holes ought to be greatly reduced with
respect to the Bondi rates estimated from the temperature and density
of the hot ISM. One way we can accommodate the lower accretion
rates is by including strong mass loss/winds in the hot accretion
flows. This is equivalent to adopting density profiles in the flows
which are flatter than those for a pure inflow ADAF; with $\rho
\propto r^{-3/2 +p}$ (see \S 4 and Fig.2). The presence of outflows
(as proposed by Blandford \& Begelman 1999) is strongly supported by
numerical simulations (Stone et al. 1999; Igumenschev et al. 1999;
2000) showing that mass loss can be a dynamical consequence of
accretion occurring in these regimes. Recent work also emphasizes that
convection (when $\alpha \approxlt 0.1$) may also lead to
significantly suppressed densities in the inner regions of hot flows
(Narayan et al. 2000; Quataert \& Gruzinov 2000) with almost no
accretion onto the black holes and $\rho \propto r^{-1/2}$. In
addition to the above explanations, which are based on the possible
consequences of the internal dynamics of a hot/quasi-spherical
accretion flow, there may also be processes that simply change the
physical conditions in the ISM in regions close to the accretion radii
of the systems such that small amounts of material are fed to the
flows (\S5.1). Any such process would need to reduce $\mdot$ by a
factor $\sim 100$ with respect to the estimated Bondi rate (Section 4,
Fig.~2).

We have also shown that given the observed low radio flux densities
(and given the resolution of the VLA) even the high-frequency radio
emission from these nuclei can be easily reproduced by standard models
of self-absorbed non-thermal synchrotron emission from the small scale
radio jets observed in these systems (Fig.~3). This implies that the
derived $\mdot$ (and/or $p$ values in the case of winds) can only
provide upper limits to the possible contribution from hot accretion
flows and that the relevant accretion rates could feasibly be
lower. The radio emission could be produced in localized regions in
the outflows/jets where the radiative efficiencies are much higher. If
$\mdot$ is small enough, a hot accretion flow may not even be required
and accretion could for example occur via a standard accretion disk
with high radiative efficiency with extremely low accretion rates (to
satisfy $L \sim 0.1 \Mdot c^2$) \footnote{A pure Bondi accretion flow
would also have even lower radiative efficiencies than an ADAF and
therefore may also be relevant, although it would be hard to
reconcile the absence of any angular momentum in the flow and the
presence of jets in these systems.}

It is important to note that the standard ADAF model is also
inconsistent with the recent detection of linear polarization (Aitken
et al. 2000) at 1mm in Sgr A$^*$. A standard ADAF model is unpolarized
at these frequencies.  In accordance with the constraints we have
obtained here for the elliptical galaxy cores, in Sgr A$^*$ the
accretion rate has to be much lower then that expected from Bondi
estimates (Agol 2000; Quataert \& Gruzinov 2000).

We propose that the small-scale radio jets present in these systems
may heat the gas in the ISM close to the accretion radii of these
systems and stifle the accretion.

\subsection{Low $\mdot$: Heating at the accretion radius by outflows/jets}

We suggest that outflows may stifle accretion by reducing $\Mdot$.
The small scales radio jets, present in all of these systems (which
may or may not be dynamically coupled to a hot accretion flow), are
likely to transfer momentum and energy to the ambient gas. Because the
sizes of the radio jets are similar to those inferred for the
accretion radii of these systems, (see \S 3; for the black hole masses
of $\sim 10^9 \Msun$) the radio jets may heat the ISM gas in that
region. If gas outside and around the accretion radius, $R_{\rm A}$,
is heated, the accretion radius decreases (potentially up to a point
where the accretion radius may not even exist) leading to a decrease
in the accretion rate (or luminosity), where $\mdot \propto R_{\rm
A}^2
\rho(R_{\rm A}) c_s(R_{\rm A}) \propto T^{-5/2}$, for a given external
pressure (see Di Matteo \& Fabian 2000 for details). Given that small
scale radio jets are observed in basically all early type galaxy
cores, we may expect the accretion rates to be reduced in most
cases. The heating of the ISM by jets/outflows may lead to similar
effects as those due to conduction (Gruzinov 1999), possibly induced
by the presence of high magnetic fields which build up in the central
regions of the hot, cooling ISM.

We note that if the accretion rate is indeed regulated by jet activity
we should expect these sources to undergo cycles of activity.  As the
jets heat up the ISM we expect the the accretion rates to decrease,
but such a decrease will most likely lead to a decline of the jet
activity itself. As the heating of the ISM is then suppressed one
might expect the fueling onto the central object to be resumed.  With
the increased accretion rates and luminosisties the cycle can be
started again.

\subsubsection{Jets and ISM pressure}

Although at lower frequencies (i.e.~at larger scales) the thermal
pressure of the hot X-ray gas (often forming a cooling flow) may be
enough to confine the radio sources (e.g. Fabbiano et a.~1987;
1988) the core regions of the jet are likely to be overpressured with
respect to the X-ray gas (see e.g.~Fig.1, where the larger scale
emission seems to have been disrupted by the cooling ISM gas at low
frequencies and only expanding buoyantly through the
medium)\footnote{This may also suggest that these sources may be
relatively young, also suggested by their fairly peaked spectral
distribution}.~Taking the estimates of $B$ from \S 4, we expect
$P_{\rm B} \sim $ a few$ \times 10^{-10}
\dynpcmsq$ in the jets. The pressure in the X-ray gas, $P_{\rm x} 
\sim n kT \sim $ a few $ 10^{-11} \dynpcmsq$ for central gas densities 
$n(ISM) \sim 0.1-0.4$ and typical temperatures $kT \sim 1 \keV$
(DM00). Simple estimates of the energetics of the core also suggest
that significant power from the radio source has been deposited in the
inner X-ray gas. The rate of energy supplied to the ISM is greater
than just the radio luminosity,  $L_{\rm radio} \sim
\nu_{\rm max} L_{\nu_{\rm max}} \sim 10^{38-39} \ergps$, and is at least
$L_{\rm j} \sim L_{\rm radio} (D/r) (c/v_{\rm j}) \sim 100 L_{\rm
radio} $ where $D/r \approxgt 10$, is the ratio of jet distance from
the central object ($D$) to its radius $r$ and $v_{\rm j} /c$, the
ratio of the jet to light speed, is typically $\approxlt 0.1$ for
these weak sources. The bolometric luminosity of the X-ray gas within
$\sim 1 \kpc $ (which is often greater than the extent of the jets) in
the galaxies, $L_{x} \sim 10^{40-41} \ergps$~(e.g.,~Allen et al.~2000)
is $\sim L_{\rm j}$, suggesting that the jet power has a strong effect
on the thermal balance of the X-ray gas.  Note that although it is
unclear whether the jets can heat most of the gas (and not just as
small fraction of it) near the accretion radii of the systems, if the
jets are overpressured with respect to the ISM gas we expect that they
will be widening sideways as they drive the shock into the surrounding
gas and may therefore affect a large area (Begelman \& Cioffi 1990).

\subsection{Suppressed accretion rates or mass loss}

In both of the models discussed in \S 4 the accretion rates onto the
central black holes in early-type galaxies must be reduced with
respect to the Bondi estimates. The low radiative efficiencies of
ADAFs are not enough to explain their low-luminosities. If mass loss is
important, the accretion rates may only be reduced in the inner regions
of the flows and the material may still be fed at the expected Bondi
rates in the outer regions, for a fixed temperature and density of the
ISM. If instead the ISM is heated near the Bondi radius the accretion
rate may be suppressed at large radii and $\mdot \ll \mdot_{\rm
Bondi}$ throughout the flows.

Follow-up X-ray observations will provide means for distinguishing
between these two possibilities. As discussed in detail in DM00,
models with strong mass loss (winds) predict significant X-ray fluxes
due to bremsstrahlung emission. Most of the contribution to the
bremsstrahlung luminosity comes from the outer regions of the flows
where the accretion rate is high and the densities are also relatively
high (see e.g. Figure 2 in DM00). If $\mdot$ is suppressed because
jets/outflows heat the ISM significantly in the outer regions of the
flow, we expect the bremsstrahlung emission, which is $\propto
\rho^2$, to be strongly reduced by a factor $\sim (\mdot/\mdot_{\rm
Bondi})^2$. Note that because of the low $L/L_{\rm Edd}$ ratios in
these sources ($\approxlt 10{-5 -6}$), even if accretion was to occur
via a standard thin disk the dominant X-ray emission mechanism would
most likely be bremsstrahlung.

We have argued that the $2-10$ \keV~ASCA hard X-ray power laws
detected in a number of nearby ellipticals (Allen et al. 2000), if
indeed produced by accretion around their central black holes, favors
strong outflow models (DM00). Observations with the Chandra X-ray
observatory are needed to clearly resolve the central sources in
these objects.  At present, because of the lack of a clear detection
of nuclear X-ray emission, it is not possible to exclude the
possibility that the accretion rates in these systems are simply much
lower than those implied the Bondi estimates.

Higher resolution VLBA observations at high radio frequencies for
$M_{BH} \sim 10^{9} \Msun$ should resolve scales as small as a few
tens of $R_{S}$ and would therefore provide a crucial test for the
presence of hot accretion flows in these systems.  Results from 15 GHz
VLBA imaging (Falcke et al.~1999) of a sample of galaxies (both
ellipticals and spirals) show that many radio cores may still be of
non-thermal origin. Further observations at higher frequencies are
necessary to clearly determine the spectrum of the core emission.

The relationship between the high-frequency radio flux and the
estimated black hole masses can also provide some clues for explaining
the quiescence of these systems. Franceschini et al.~(1998) have shown
that there is a direct relation between the 5 GHz radio cores in
ellipticals and their central black hole masses, which may be accounted
for by the thermal synchrotron emission from an ADAF. If such a proposal
is valid and a low-$\mdot$ ADAF (Fig.~2) is indeed producing most of
the emission, the relationship should hold at higher frequencies. In
the next section we derive the expected relationship when
outflows are present or when heating of the ISM is occurring.

\subsection{Radio power and black hole masses}

The high frequency radio observations have allowed both the
synchrotron flux and the position of the peak, if present, to be
measured.  In Figure 4 we plot the radio luminosity at 8 and 22 GHz
versus the measured black hole masses for the seven objects (from DM99
and the four here). Although the sample is small and only spans a
limited range of black hole masses, much smaller than that plotted in
Franceschini et al. (1998), we can still compare our results to the
relationship they have found. This allows us to see whether the
Franceschini et al. relationship holds at higher frequencies where the
contribution from a potential ADAF is most relevant. We also show that
the presence of outflows in an ADAF or of significant heating in the ISM,
would change the relationships between radio core flux and black hole
masses.

Figure 4 shows the radio core luminosities and total radio
luminosities for NGC 4649, NGC 4472, NGC 1399, NGC 2300, NGC 4594, NGC
4278 and M87 plotted against their black hole masses. We plot the
fluxes measured at 8 and 22 GHz together with the previously reported
5 GHz values (we do not plot the $43$ GHz fluxes because at this
frequency we only have firm detections for 4 objects).  The
relationship depicted from the fluxes measured at 22 GHz (solid dots -
Figure 4) suggests that the high frequency core fluxes may be more
strongly correlated to $M_{\rm BH}$. Note that the 5 GHz fluxes, would
indicate a much flatter relationship between $M_{\rm BH}$ and
$L_{\nu}$. This suggests that the relatively weak, arcsec-scale radio
jet components, which contribute more predominantly at lower
frequencies may not be as as strongly correlated in this plane.

In Figure 4 we also show the expected correlation (solid line)
between radio luminosisity and black hole mass for an ADAF (dashed
line; Franceschini et al. 1998) and for an ADAF with winds (dotted
line). This relationship is obtained by noting that the Rayleigh Jeans
part of the synchrotron spectrum scales as $L_{\nu_{\rm c}} \propto
\nu_{\rm c}^{2} R^2$ with $\nu_{\rm c} \propto B \propto R^{-5/4}
\Mdot^{1/2}M_{\rm BH}^{1/4}$. For $\Mdot \propto
\Mdot_{\rm Bondi} R^p$ we have
\begin{equation}
R \propto  \nu_{\rm c}^{4/(-5+2p)} \Mdot_{\rm Bondi}^{2/(5-2p)} M_{\rm BH}^{1/(5-2p)} 
\end{equation}
and given that $\Mdot_{\rm Bondi} \propto M_{\rm BH}^2 n(ISM)/c_{\rm
S}^3(ISM)$ and following Franceschini et al. (1998) taking $n\propto M_{\rm
gal}$ and $c_{\rm s} \propto M_{\rm BH}^{1/4}$ with $M_{\rm gal}
\propto M_{\rm BH}$ for ellipticals (e.g.~Magorrian et al.~1998), we
find, using Eq.~(4),
\begin{equation}
L_{\nu} \propto \nu_{\rm c}^{(4p-2)/(2p-5)} M_{\rm BH}^{11/(5-2p)};
\end{equation}
which recovers $L_{\nu} \propto \nu^{2/5} M_{\rm BH}^{2.2}$ derived by
Franceschini et al. for the case of $p=0$. Relation (5) implies that
the presence of outflows i.e. for $0 < p \approxlt 1$ steepens the
relationship between core power and black hole mass.  The solid line
in Figure 4 is the steep correlation $L_{\nu} \propto M_{\rm
BH}^{3.6}$ obtained for $p=1$. A few of the objects seem in better
agreement with the steeper relationship, although a larger number of
objects would be needed to distinguish any clear trend.  Note also
that in their regression analysis for 8 objects, Franceschini et
al. found $L_{\nu} \propto M_{\rm BH}^{2.66-2.73}$ for the total and
core radio luminosities respectively.

We note that Franceschini et al. (1998) also plot objects like M31 and
Sgr A$^*$ and their relationship extends down to $M_{BH} \sim 10^6
\Msun$. Here we could also plot higher frequency radio measurements of
those objects (and e.g. NGC 4258) but for those systems Bondi
accretion from the ISM would not apply as indicated by Eq.~8 and the
expected relationship may be flatter with $L_{\nu} \propto \mdot^{6/5}
M_{BH}^{8/5}$. Relation (5) is only recovered where the Bondi argument
for the accretion rates can be applied.


The other possibility discussed above is that outflows stifle the
accretion by reducing $\Mdot$. This could be both due to the possibly
poorly collimated outflows from the ADAF or in any case to the
observed radio jets present in all these systems. 
In this case we predict a different relationship between
radio power and mass accretion rate.

Heating of the gas at the accretion radius implies an effective 
energy flux into the ISM and a relationship with the central 
black hole mass given by, 
\begin{equation}
L_{\rm j} \propto \Mdot T(ISM) \propto M_{\rm BH}^3/c_{\rm s}(ISM) \propto
M_{\rm BH}^{2.75}.
\end{equation} 
Given that $L_{\rm j} \propto L_{\rm radio}$ if heating by the jets is
important we would also expect a strong dependence of the total radio
power on the central black hole mass.  In Figure 5 we plot the radio
power $\nu_{max} L_{\nu_{max}}$ versus black hole mass. Although the
errors on the black hole mass estimates are not well-known and could
easily be large, Fig.~5 seems to show a more consistent trend than
Figure 4. The two most powerful radio sources in the sample, NGC 4278
and M87, are the ones that depart the most from the above relation.

~~~In summary, Figure 4 and 5 suggest that the radio power at high
radio frequencies is likely to be a good tracer of black hole mass and
useful for testing the accretion properties of these systems.The
Franceschini et al. relationship between core radio power and black
hole mass may get steeper at higher frequencies or in the presence of
outflows. 

Obtaining a large enough sample of elliptical galaxies observed at
high resolution at high radio frequencies spanning a larger range of measured
black hole masses may be useful for discriminating whether the radio
emission is due to the weak radio jets in these systems or an ADAF
component. In particular, with a statistical sample of objects one can
test for the presence of strong outflows in ADAFs, or look for signs
of interactions between jets and ISM, which can lead to suppression of
the accretion onto these systems.  Resolving the correlation between
black hole mass and radio flux can therefore break the degeneracy
between the two interpretations for the origin of the radio emission:
the case for small accretion rates throughout the flows (i.e.~the case
where $\Mdot_{\rm Bondi}$) or the situation where most of the material is
lost through a wind and small amounts are accreted onto the black
holes, $\Mdot\propto \Mdot_{\rm Bondi} R^p$.

\acknowledgements
~~~~ T.\,D.\,M.\ acknowledges support for this work provided by NASA
through Chandra Postdoctoral Fellowship grant number PF8-10005 awarded
by the Chandra Science Center, which is operated by the Smithsonian
Astrophysical Observatory for NASA under contract NAS8-39073. ACF
thanks the Royal Society for support. The Very Large Array is operated
by the National Radio Astronomy Observatory, which is a facility of
the National Science Foundation operated under cooperative agreement
with Associated Universities Inc.

~~~~~~~

\newpage
~~~
\begin{table*}[t]
{\footnotesize 
\centering
\caption[t:vla]{\label{t:vla} VLA data.}
\begin{tabular}{cccccccc}\hline
\\
Object & Black Hole & Frequency & RMS & Total & Peak & Peak &
        Position\\ \vspace{0.3cm} &&&&&&at 8GHz resolution& \\
\vspace{0.3cm}& $\Msun$ &$\nu$ (GHz)& {\rm mJy}& $F_{\nu}$({\rm mJy}) &$F_{\nu}$({\rm mJy}) & $F_{\nu}$({\rm mJy}) &  J2000 \vspace{0.3cm} \\ 
\hline
\\
NGC 1399 & $5.2 \times 10^9$ &8.4& 0.05 & $159$ & $21.6\pm0.4$ & $21.6\pm0.4$  & 03 38 29.0 -35 27 01.0\\
	
         &   &22& 0.30 & $59$& $ 22.2\pm1.2$  &  $25.4\pm1.5$& 03 38 29.0 -35 27 00.8\\
	 &  &43 & 0.7 & 19 & $16.1\pm1.6$ & $17.9\pm2.0$ & 03 38 29.0 -35 27 01.1\\   
\\
NGC 2300 & $2.7 \times 10^9$ &8.4& 0.05 & $0.76$ & $0.76\pm 0.05 $& -- & 07 32 20.2 +85 42 32.5 \\
	 &  &22 & 0.25 & 1.5 &$ 1.2\pm0.25$ & -- & 07 32 19.9 +85 42 32.6 \\
	 &  &43 & 0.4 & -- & $ < 1.3$  & -- & 07 32 20.7  +85 42 27.6\\
\\
NGC 4649 & $3.9 \times 10^9$&8.4 &0.05  & 22.7 &  $20.4\pm0.4$  & $20.4\pm0.4$ & 12 43 39.96 +11 33 09.8 \\
	 & &22  &0.25  & 20.7 & $19.7\pm1.0$  & $20.4\pm1.0$  & 12 43 39.97 +11 33 09.6 \\ 
	 & &43  &0.38  & 15.9  &  $14.5\pm1.4$  & $16.2\pm2.0$ & 12 43 39.97 +11 33 09.7\\
\\
NGC 4594 & $2 \times 10^9$ & 8.4 & 0.53 & $137\pm2.7$ & $136\pm 2.7$ & &  \\
&                          & 22  & 0.41 & $93\pm 4.6$  & $90.4\pm4.5$ && 12 39 59.4 -11 37 23.0 \\
&                          & 43  & 0.76 & $57\pm 6$   & $53\pm 5 $  && \\
\\
NGC 4278 & $1.6 \times 10^9$ & 8.4 & 0.20 & $133\pm2.6$ & $129\pm2.6$ & & \\
         &                   & 22  & 0.26 & $67\pm 3.3$  & $65\pm 3.2$  && \\
         &                   & 43  & 0.31 & $51\pm 5 $ &  $50 \pm 5$ && 12 20 06.8 29 16 50.8 \\

\hline
\\
\end{tabular}
}
\end{table*}

\newpage

\vbox{\centerline{
\vbox{
\psfig{figure=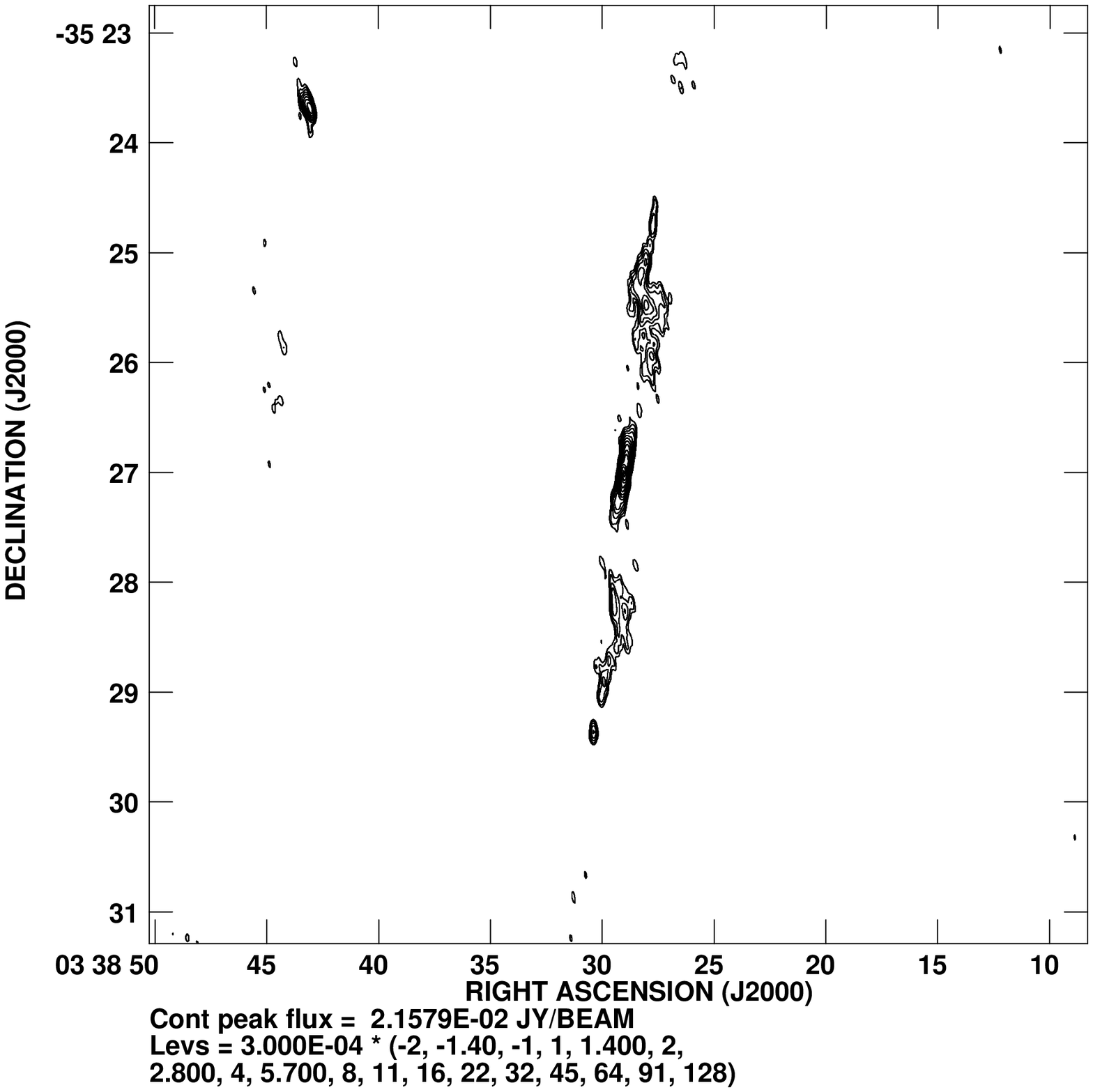,width=0.5\textwidth,angle=0}
\psfig{figure=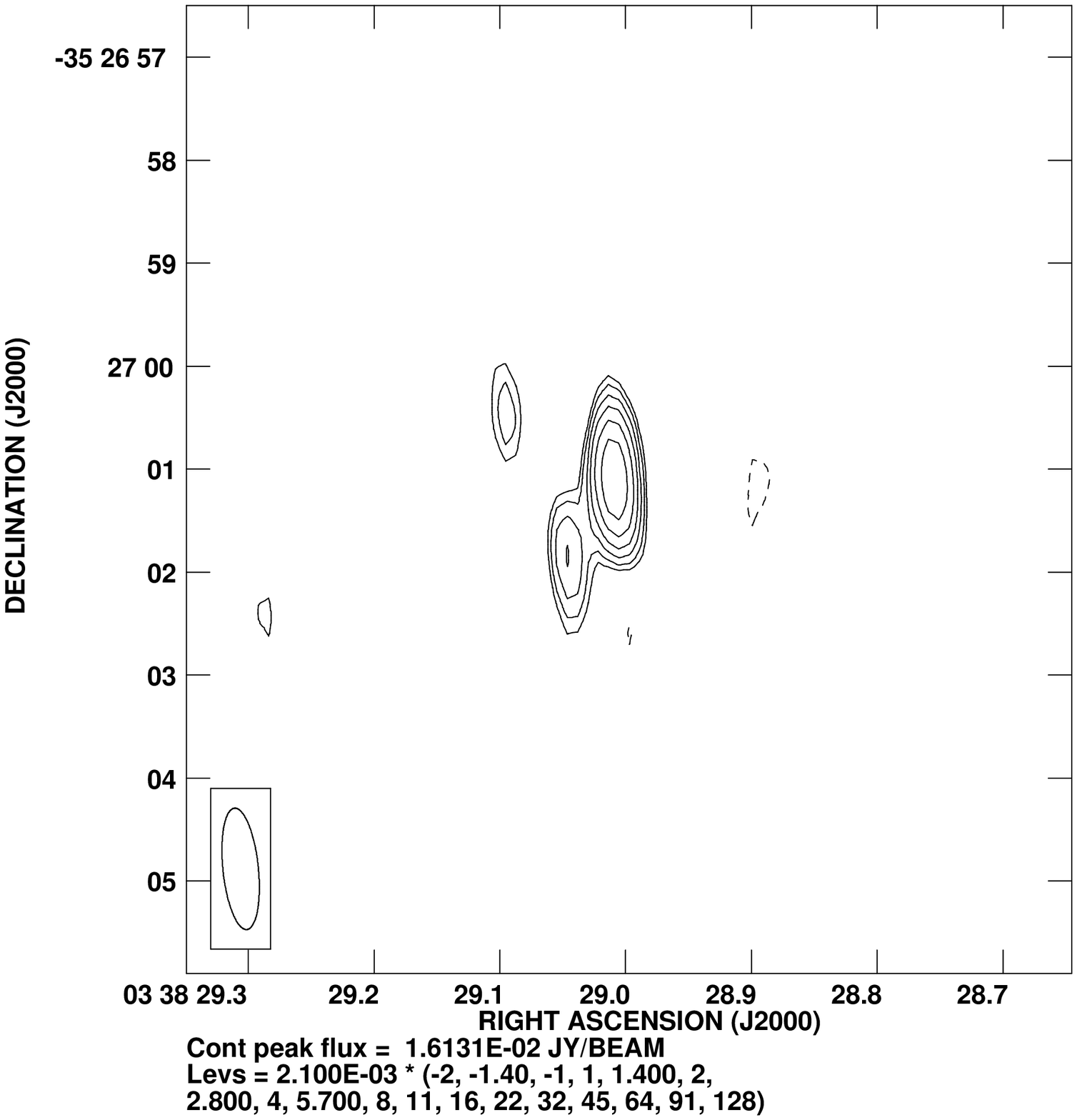,width=0.5\textwidth}}}
\figcaption[]{\footnotesize VLA high resolution radio images for NGC 1399 at 8 and 43 GHz. The FWHM of the restoring Gaussian beam is shown in the box in lower left corner.}}

\newpage

\vbox{\centerline{
\vbox{
\psfig{figure=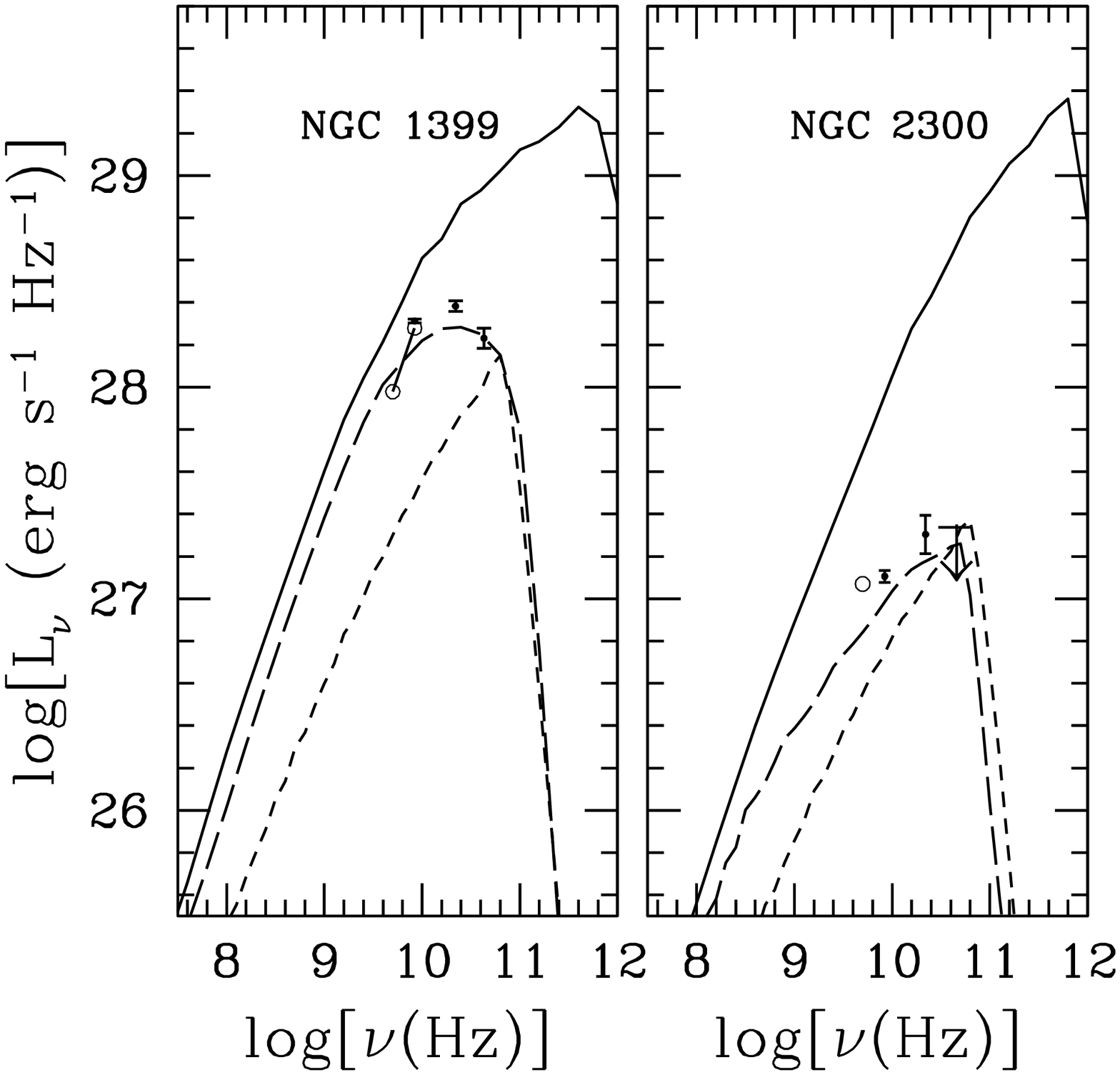,width=0.5\textwidth}
\vspace{-0.2cm}
\psfig{figure=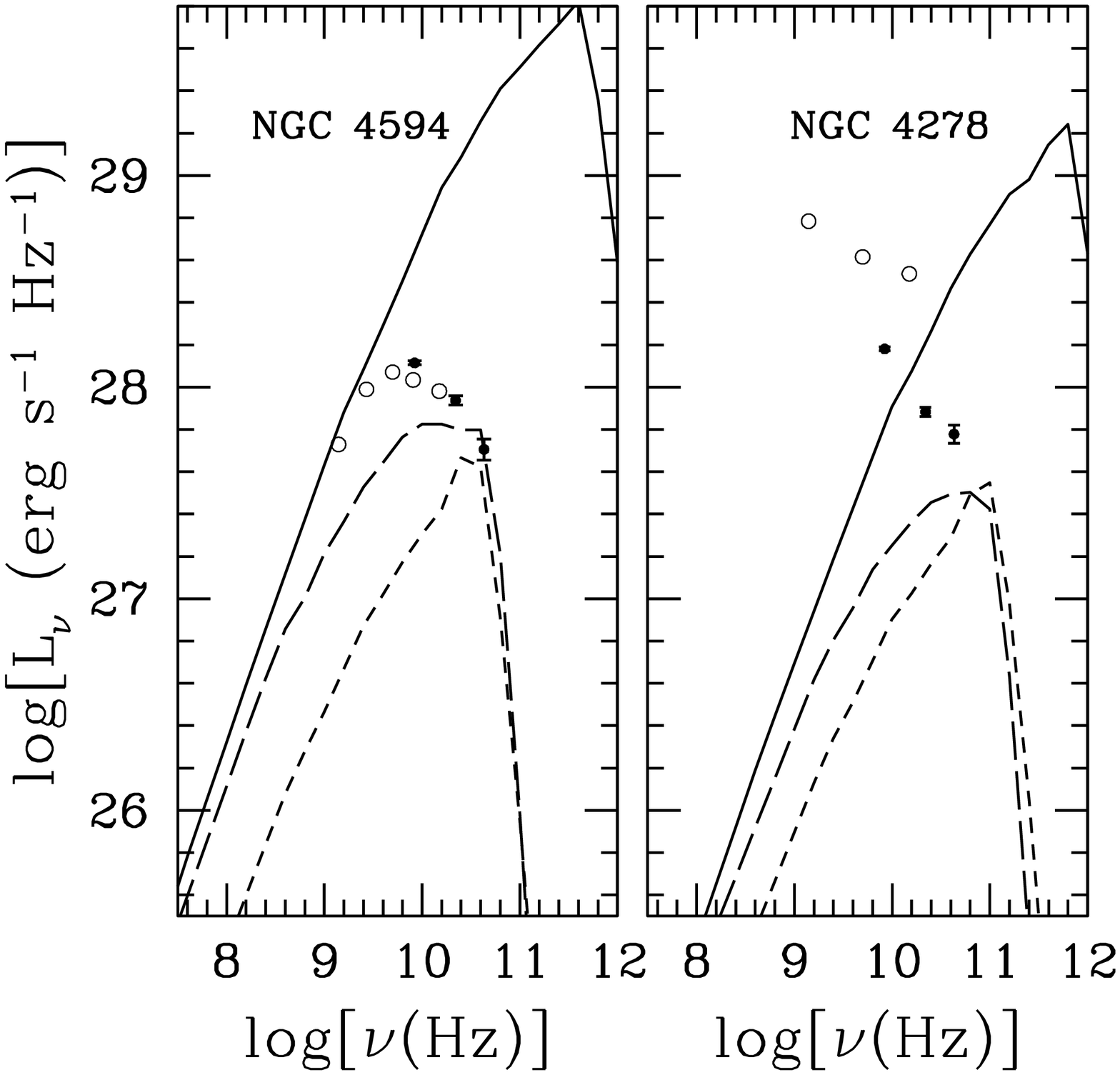,width=0.5\textwidth}}}

\figcaption[]{\footnotesize The thermal synchrotron emission spectra from
hot accretion flows with and without outflows. The solid dots are the
VLA fluxes from Table 1. The open circle measurements obtained from
literature. The solid line shows pure inflow ADAF models with $\mdot
\sim \mdot_{\rm Bondi}$, the long dashed lines models with strong
outflows $\mdot \propto \mdot_{\rm Bondi} r^p$ and the short dashed
lines ADAF models without outflows but with $\mdot \ll \mdot_{\rm
Bondi}$.}  }
\vspace*{0.5cm}

\newpage

\vbox{
\centerline{
\psfig{figure=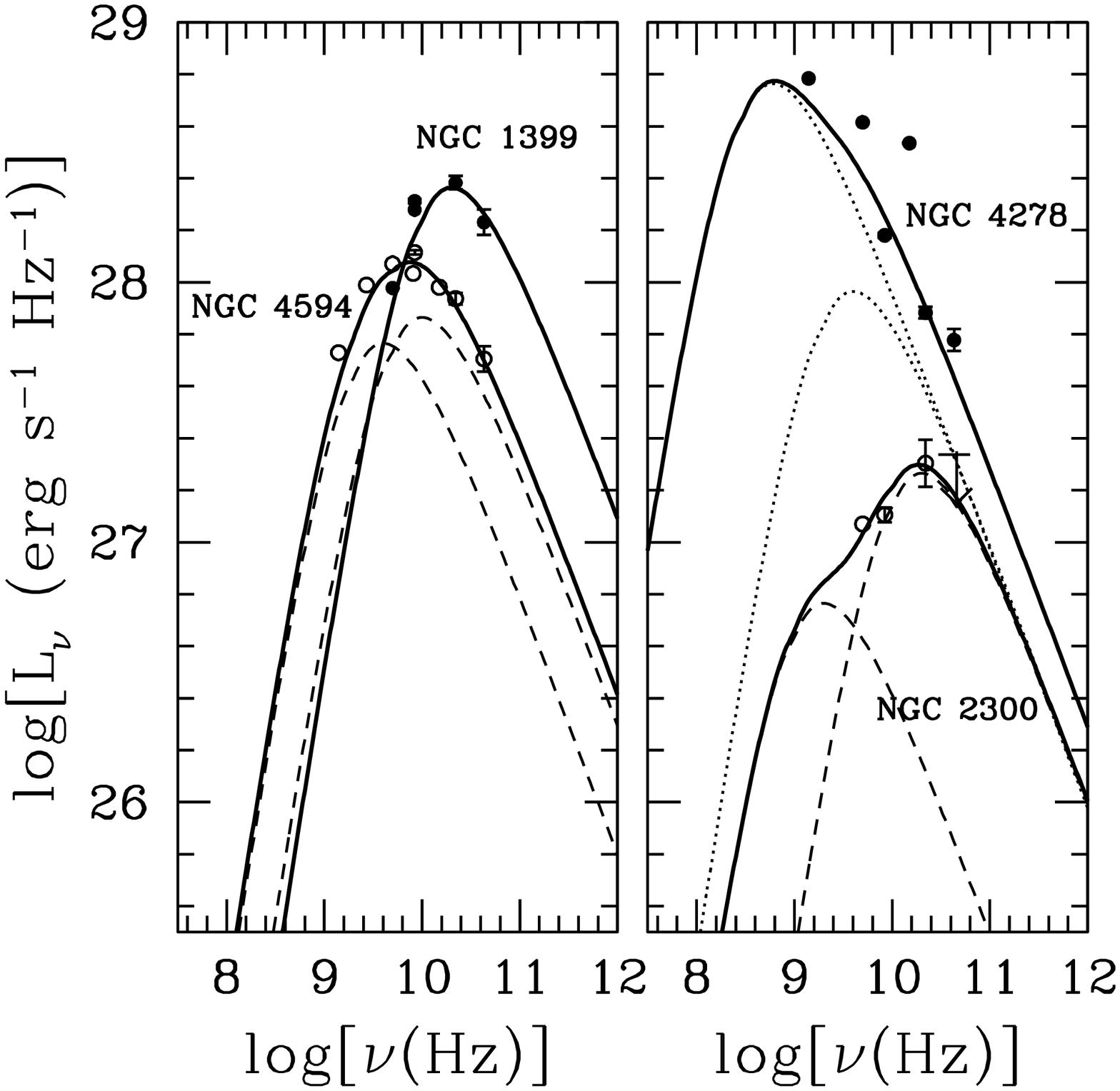,width=0.8\textwidth}}
\figcaption[]{\footnotesize Spectra made by superposition of non-thermal
self-absorbed synchrotron regions. The solid lines are the sum of the
various components the dashed lines and dotted lines show the separate
components. At these frequencies NGC 4594 can be well-explained with
only one region. For the other objects two-regions are modeled. All
models have $\xi =3$. The data points are the same as in Figure 2.} }
\vspace*{0.5cm}

\newpage

\vbox{ \centerline{
\psfig{figure=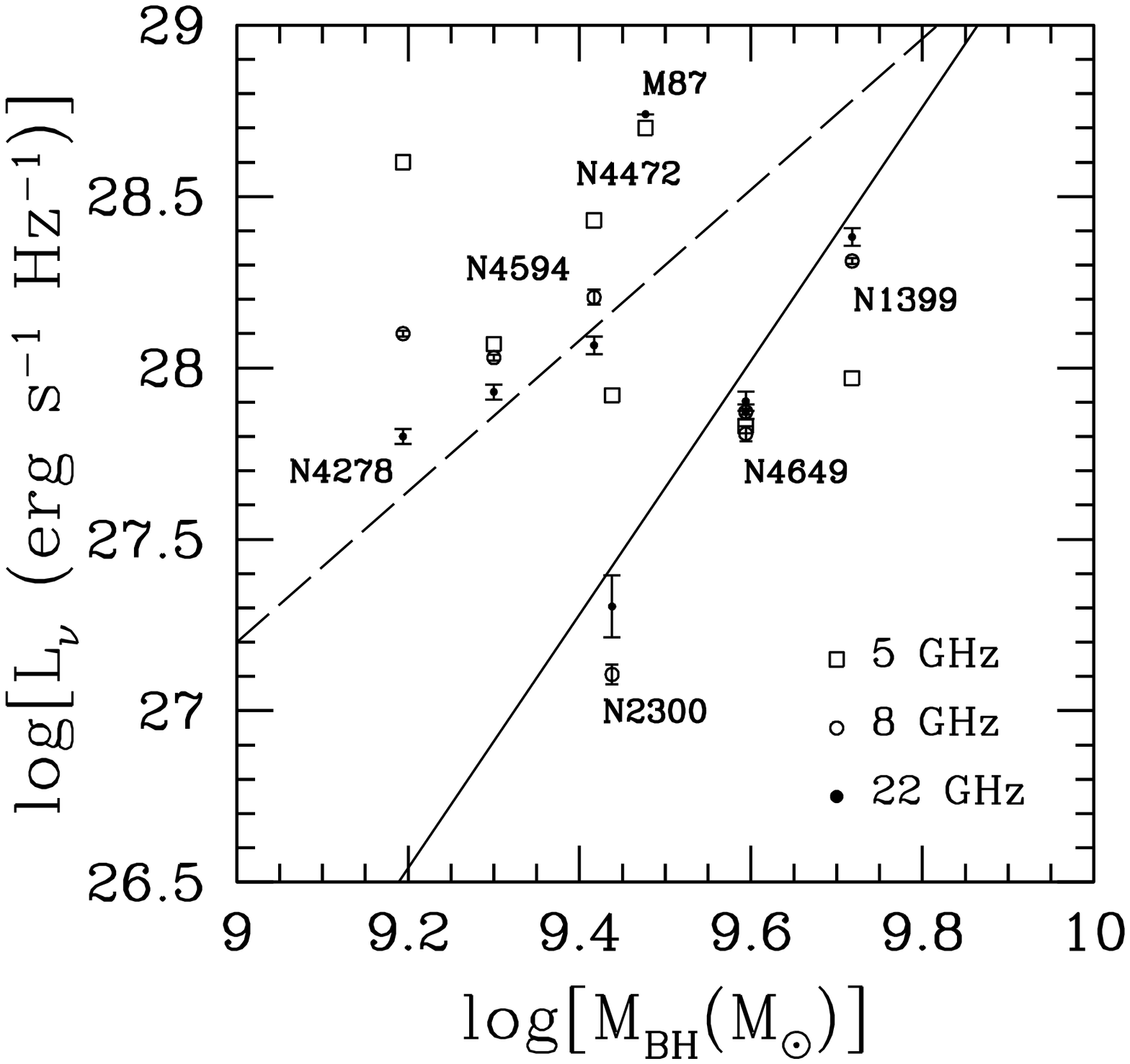,width=0.8\textwidth}
}}
\figcaption[]{\footnotesize The core radio power
at different frequencies. The dashed lines correspond to the ADAF based relationship $L \propto M_{BH}^{2.2}$ the solid line to ADAF + winds with $L \propto
M_{BH}^{3.6}$. }
\vspace*{0.5cm}

\newpage
\vbox{ \centerline{
\psfig{figure=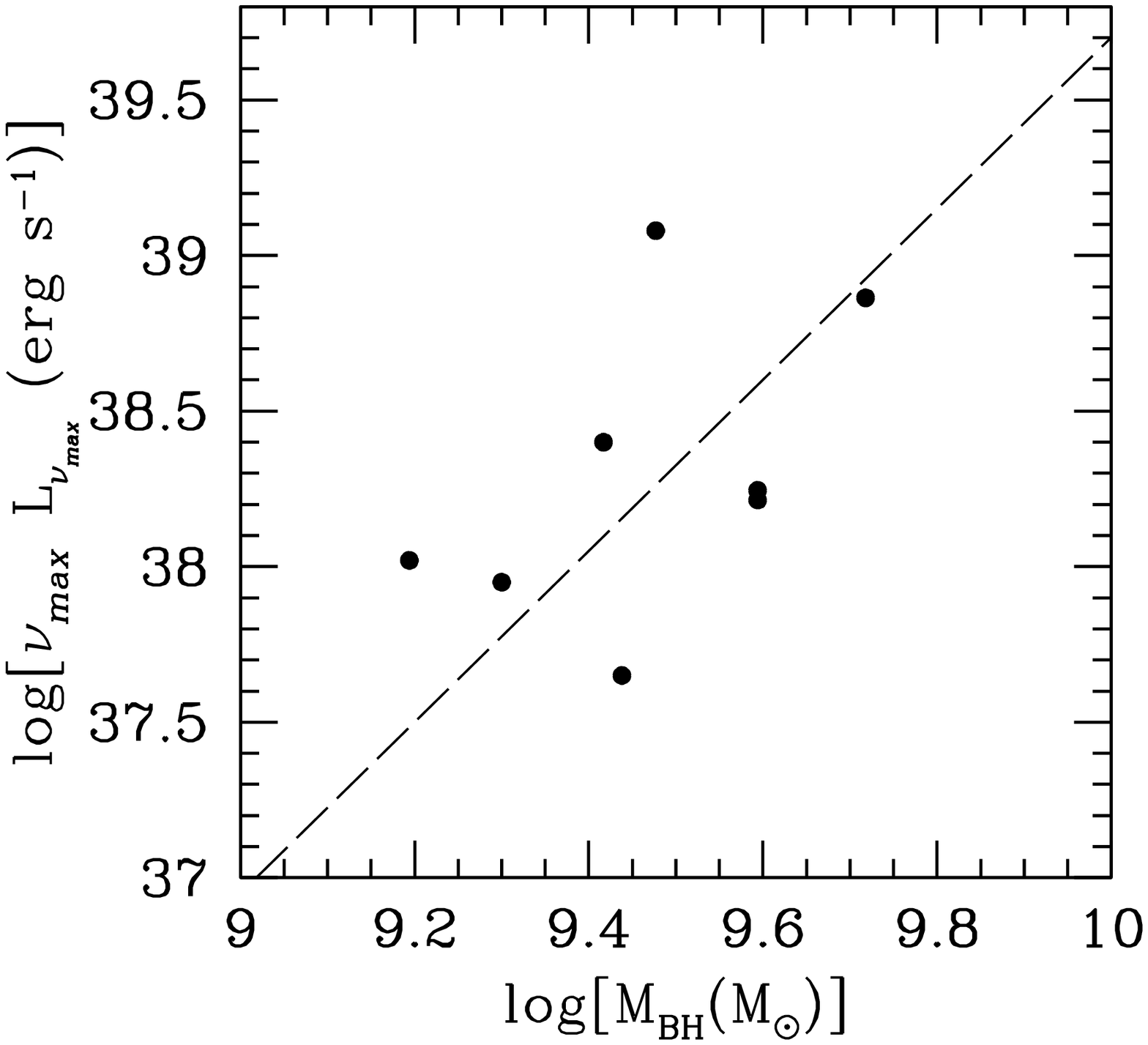,width=0.8\textwidth}
}}
\figcaption[]{\footnotesize The peak radio luminosity as a 
function of black hole mass. The same objects as in Figure 4 are plotted.
The solid lines is the relationship expected if the radio jets heat the ISM.}

\vspace*{0.5cm}

\end{document}